\documentclass[%
aps,
pra, 
reprint,
 amsmath,amssymb,
 aps,superscriptaddress,
]{revtex4-1}

\usepackage{graphicx}
\usepackage{dcolumn}
\usepackage{bm}
\usepackage{amsmath}
\usepackage[multiple]{footmisc}

\def\mydash{{\hbox{-}}}
\newcommand{\func}{}

\begin{document}

\title{Negative Kerr nonlinearity of graphene \\ as seen via chirped-pulse-pumped self-phase modulation}

\author{Nathalie Vermeulen}
\affiliation{Brussels Photonics Team, Dept. Applied Physics and Photonics, Vrije Universiteit Brussel, Pleinlaan 2, 1050 Brussel, Belgium}
\author{David Castell\'o-Lurbe}
\affiliation{Brussels Photonics Team, Dept. Applied Physics and Photonics, Vrije Universiteit Brussel, Pleinlaan 2, 1050 Brussel, Belgium}
\author{JinLuo Cheng}
\affiliation{Brussels Photonics Team, Dept. Applied Physics and Photonics, Vrije Universiteit Brussel, Pleinlaan 2, 1050 Brussel, Belgium}
\affiliation{Department of Physics, University of Toronto, 60 St. George St., Toronto ON,
M5S 1A7, Canada}
\author{Iwona Pasternak} 
\affiliation{Institute of Electronic Materials Technology, Wolczynska 133, 01-919 Warsaw, Poland}%
\author{Aleksandra Krajewska} 
\affiliation{Institute of Electronic Materials Technology, Wolczynska 133, 01-919 Warsaw, Poland}%
\author{Tymoteusz Ciuk} 
\affiliation{Institute of Electronic Materials Technology, Wolczynska 133, 01-919 Warsaw, Poland}%
\author{Wlodek Strupinski} 
\affiliation{Institute of Electronic Materials Technology, Wolczynska 133, 01-919 Warsaw, Poland}%
\author{Hugo Thienpont}
\affiliation{Brussels Photonics Team, Dept. Applied Physics and Photonics, Vrije Universiteit Brussel, Pleinlaan 2, 1050 Brussel, Belgium}
\author{J$\mathrm{\ddot{u}}$rgen Van Erps}
\affiliation{Brussels Photonics Team, Dept. Applied Physics and Photonics, Vrije Universiteit Brussel, Pleinlaan 2, 1050 Brussel, Belgium}


\begin{abstract}
We experimentally demonstrate a negative Kerr nonlinearity for quasi-undoped graphene. Hereto, we introduce the method of chirped-pulse-pumped self-phase modulation and apply it to graphene-covered silicon waveguides at telecom wavelengths. The extracted Kerr-nonlinear index for graphene equals $n_{2,\mathrm{gr}} = - 10^{-13}$~m$^2$/W. Whereas the sign of $n_{2,\mathrm{gr}}$ turns out to be negative in contrast to what has been assumed so far, its magnitude is in correspondence with that observed in earlier experiments. Graphene's negative Kerr nonlinearity strongly impacts how graphene should be exploited for enhancing the nonlinear response of photonic (integrated) devices exhibiting a positive nonlinearity. It also opens up the possibility of using graphene to annihilate unwanted nonlinear effects in such devices, to develop unexplored approaches for establishing Kerr processes, and to extend the scope of the ``periodic poling'' method often used for second-order nonlinearities towards third-order Kerr processes. Because of the generic nature of the chirped-pulse-pumped self-phase modulation method, it will allow fully characterizing the Kerr nonlinearity of essentially any novel (2D) material.         
\end{abstract}



\maketitle

\section{Introduction}

The emergence of graphene, a 2D honeycomb lattice of carbon atoms with a linear electronic band structure, has had an enormous impact on numerous research areas, including that of nonlinear optics. Already in 2010 Hendry and co-workers found the Kerr nonlinearity of graphene \footnote{We here refer to the parametric Kerr effect, which differs from the non-parametric saturable absorption effect in graphene} to be extremely strong -- a direct result from the material's unique band structure -- \cite{Hendry}, and this experimental observation attracted much attention from the broad community working on advanced Kerr-nonlinear light generation phenomena such as supercontinuum and frequency-comb generation \cite{Holzwarth, Kippenberg, Savage}. To extract graphene's Kerr nonlinearity, Hendry \emph{et al.} used ``free-space'' laser excitation to establish four-wave mixing (FWM) in a graphene sample, and later on also free-space Z-scan measurements \cite{Kockaert} were employed for this purpose. Meanwhile, the ever-increasing importance of integrated photonics encouraged researchers to combine graphene's nonlinear optical properties with on-chip (semiconductor) waveguide structures \cite{Gu, Zhou, Ji, Hu}. Covering a waveguide with graphene dramatically changes the waveguide's nonlinear behavior because of the interaction between the evanescent tails of the guided mode and the 2D top layer. Despite graphene's very strong Kerr nonlinearity as was already demonstrated in 2010, a rather limited number of on-chip nonlinear-optical graphene devices have been reported on so far, and they mostly rely on FWM in graphene-covered silicon waveguides \cite{Gu, Zhou, Ji, Hu}. To pave the way to a more varied exploitation of graphene in nonlinear photonic integrated circuits and enable advanced nonlinear light generation, not only the strength or magnitude of graphene's Kerr nonlinearity has to be known but also its sign needs to be measured, since the latter determines whether the nonlinear effects induced by the graphene and the underlying waveguide add or subtract. The Kerr coefficient of the graphene cover layer has been (tacitly) assumed to be positive in the demonstrated FWM devices, but in fact FWM intrinsically only allows measuring the coefficient's magnitude and not its sign. If with another method one can also assess the sign, this knowledge will enable researchers to fully exploit graphene's promise as efficiency-enhancing material in essentially any nonlinear photonic system. In this article, we examine both the magnitude and the sign of graphene's Kerr nonlinearity using a different approach, namely by means of chirped-pulse-pumped self-phase modulation (SPM) experiments in graphene-covered silicon waveguides.

\section{Basic principles of chirped-pulse-pumped SPM}     

The following explanation on the physics of the chirped-pulse-pumped SPM approach shows that, as opposed to FWM experiments, it does enable assessing the sign of a material's Kerr nonlinearity. SPM is a Kerr process that broadens or narrows the spectral width of an input laser pulse \cite{Agrawal, Pinault}. This broadening or narrowing effect can become very pronounced even at modest excitation powers, provided that the input pulse exhibits chirp. Let us consider a waveguide section of length $\Delta z$ in which a pulse propagates with amplitude $A(z,t)=|A(z,t)|\exp\left[i\varphi(z,t)\right]$. For a Fourier-transformed amplitude $\tilde{A}(z, \omega-\omega_0)$ centered at frequency $\omega_0$, the square of the $rms$ spectral width can be written as $\mu_2(z) = (\int_{-\infty}^{\infty}(\omega-\omega_0)^2|\tilde{A}(\omega-\omega_0)|^2\mathrm{d}\omega)/(\int_{-\infty}^{\infty}|\tilde{A}(\omega-\omega_0)|^2\mathrm{d}\omega)$ \cite{Pinault, CastelloLurbe}, which in the time domain can be expressed as $\mu_2(z)=(\int_{-\infty}^{\infty}|\partial_t A|^2\mathrm{d}t)/(\int_{-\infty}^{\infty}|A|^2\mathrm{d}t)$ \cite{Pinault}. For the experiments reported in this Letter, the pulse propagation is mostly determined by Kerr-nonlinear effects rather than dispersion and free-carrier effects (which was also verified numerically with the inclusion of graphene-generated free carriers \cite{CastelloLurbe, Vermeulen}). Hence, we can adopt a simplified version of the nonlinear Schr$\mathrm{\ddot{o}}$dinger equation incorporating only SPM effects and linear loss: $\partial_z A = -(\alpha/2)A + i\gamma_{\!_K}|A|^2 A$, with $\gamma_{\!_K}$ the Kerr nonlinearity and $\alpha$ the linear loss coefficient. This yields:

\begin{align}
|A(\Delta z,t)|^2=\exp[-\alpha\Delta z]|A(0,t)|^2 && \nonumber \\ 
\varphi(\Delta z,t) = \varphi(0,t) + \gamma_{\!_K}|A(0,t)|^2{\Delta z}_\mathrm{eff} 
\end{align}

\noindent  
where ${\Delta z}_\mathrm{eff} = \left[1-\exp(-\alpha\,\Delta z)\right]/\alpha$ represents the effective length. When rewriting $A(0,t)$ as $A(0,t)=\sqrt{P_0}U_0(t)\exp\left[i\varphi_0(t)\right]$ with $P_0$ the input peak power, implementing Eq.~(1) in $\mu_2(z)$ results in the following expression for the relative spectral broadening factor $\mu_2(\Delta z)/\mu_2(0)$ after propagation over $\Delta z$:

\begin{equation}
\frac{\mu_2(\Delta z)}{\mu_2(0)}=1+\frac{2\gamma_{\!_K} P_0 {\Delta z}_\mathrm{eff}\displaystyle{\sigma_{3}/\sigma_1}+
(\gamma_{\!_K} P_0 {\Delta z}_\mathrm{eff})^2\displaystyle{\sigma_{4}/\sigma_1}}{1+\displaystyle{\sigma_{2}/\sigma_1}}
\end{equation}

\noindent
with $\sigma_{1,2,3,4}$ being the input pulse shape factors:

\begin{align}
\sigma_1 = \frac{\int_{-\infty}^{\infty} (\partial_t U_0)^2\mathrm{d}t}{\int_{-\infty}^{\infty} U_0^2\mathrm{d}t}
&& \sigma_2 = \frac{\int_{-\infty}^{\infty} (\partial_t \varphi_0)^2 U_0^2\mathrm{d}t}{\int_{-\infty}^{\infty} U_0^2\mathrm{d}t}
&& \nonumber \\
\sigma_3 = \frac{\int_{-\infty}^{\infty} (\partial_t\varphi_0) (\partial_t U_0^2) U_0^2\mathrm{d}t}{\int_{-\infty}^{\infty} U_0^2\mathrm{d}t}
&& \sigma_4 = \frac{\int_{-\infty}^{\infty} (\partial_t U_0^2)^2 U_0^2\mathrm{d}t}{\int_{-\infty}^{\infty} U_0^2\mathrm{d}t}.
\end{align}

\noindent
When assuming a quadratic input phase profile $\varphi(0,t) = -C_0\displaystyle{t^2/T_0^2}$ with input chirp parameter $C_0$, Eq.~(2) yields:

\begin{equation}
\frac{\mu_2(\Delta z)}{\mu_2(0)}=1+\frac{2\gamma_{\!_K} P_0 {\Delta z}_\mathrm{eff} C_0 \sigma_\mathrm{31}+
(\gamma_{\!_K} P_0 {\Delta z}_\mathrm{eff})^2\sigma_\mathrm{41}}{1+4\,C_0^2\,\sigma_\mathrm{21}}
\end{equation}

\noindent
with the new dimensionless shape factors $\sigma_{21,31,41}$ all having positive values determined by:

\begin{align}
\sigma_{21}=\frac{1}{T_0^4}\frac{\int_{-\infty}^{\infty}t^2\,U_0^2\mathrm{d}t}{\int_{-\infty}^{\infty}(\partial_t U_0)^2\mathrm{d}t}
&& \sigma_{31}=\frac{1}{T_0^2}\frac{\int_{-\infty}^{\infty}U_0^4\mathrm{d}t}{\int_{-\infty}^{\infty}(\partial_t U_0)^2\mathrm{d}t}
&& \nonumber \\
\sigma_{41}=\frac{\int_{-\infty}^{\infty}(\partial_t U_0^2)^2\,U_0^2\mathrm{d}t}{\int_{-\infty}^{\infty}(\partial_t U_0)^2\mathrm{d}t}.
\end{align} 

\noindent
For a Gaussian pulse with $U_0^2 =$ exp$(-t^2/T_0^2)$ and a sech$^2$ pulse with $U_0^2 =$ sech$^2(t/T_0)$, these shape factors become, respectively: $\sigma_{21} = 1$ and $\pi^2/4$; $\sigma_{31} = \sqrt{2}$ and $2$; $\sigma_{41} = 4/(3 \sqrt{3})$ and $32/35$. We point out that for $C_0 = 0$ only the term containing $\sigma_{41}$ remains, and the formula reported in \cite{Pinault} is recovered.

Eq.~(4) in fact corresponds to earlier presented SPM formulas \cite{Agrawal, Pinault}, while also taking into account the input chirp parameter. In the high-power regime, the term proportional to $(\gamma_{\!_K} P_0 {\Delta z}_\mathrm{eff})^2$ in Eq.~(4) will be dominant, and the broadening factor will be larger than 1 with the highest value attained at $C_0=0$, i.e. for a transform-limited input pulse \cite{Boyraz}. At low powers, however, the term proportional to $2\gamma_{\!_K} P_0 {\Delta z}_\mathrm{eff} C_0$ becomes important, and depending on the sign of $\gamma_{\!_K} C_0$ \footnote{We point out that for \emph{temporal} pulse broadening or narrowing in a waveguide with dispersion $\beta_2$, the sign of $\beta_2 C_0$ plays an analogous role \cite{Agrawal}}, the broadening factor can either be larger or smaller than 1 (i.e.~one has either broadening or narrowing), with a maximum or minimum value at a non-zero chirp. As such, when exciting a waveguide with a moderate-power pulse exhibiting an appropriately chosen chirp $C_0$, the nature of the spectral width modification, i.e.~broadening or narrowing, will indicate the sign of the waveguide's Kerr nonlinearity $\gamma_{\!_K}$. This is also illustrated in Fig.~1 depicting the spectral broadening factor of Eq.~(4) as a function of $C_0$ for a silicon waveguide (see further on), the Kerr nonlinearity of which is known to be positive at wavelengths above $1000$~nm \cite{Bristow}. In Fig.~1 the spectral broadening and narrowing obtained at, respectively, substantially positive and negative $C_0$-values indeed unambiguously demonstrate the positive sign of silicon's Kerr nonlinearity \footnote{In the low-power regime, spectral broadening can also be obtained at a negative input chirp (see centre region of Fig.~1), provided that the chirp value is sufficiently close to zero to let the last term in Eq.~(4) dominate.}. In a similar way, by carrying out chirped-pulse-pumped SPM experiments with graphene-covered silicon waveguides and factoring out silicon's nonlinear response, we can determine the sign of graphene's Kerr nonlinearity.  

\begin{figure}
\includegraphics[width=2.7in]{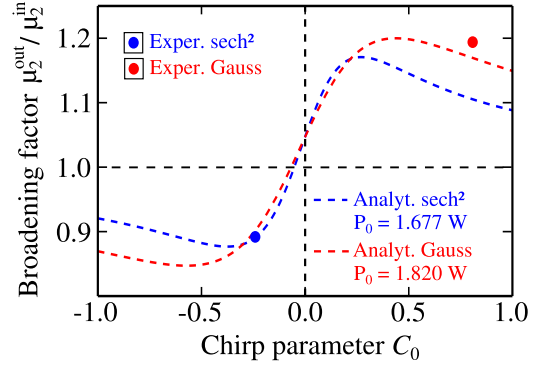}
\caption{\label{fig:BroadeningvsChirp} Illustration of the broadening factor dependence on the input chirp parameter for a silicon waveguide and for Gaussian and sech$^2$ input pulses as used in our experiments.}
\end{figure} 

\section{Experimental setup and results}

To perform such chirped-pulse-pumped SPM experiments, we employed grating-coupled bare and graphene-covered silicon-on-insulator (SOI) waveguide structures designed for operation around the $1550$~nm telecom wavelength, and all featuring the same layout for the gratings, tapers and the actual waveguide section (see Fig.~2(a)). The waveguide's cross-section has a width of $450$~nm and a height of $220$~nm which is standard for SOI chips fabricated by ePIXfab \cite{epixfab}. Because of the identical layouts of the bare and graphene-covered SOI waveguides, the bare waveguide can serve as a reference structure with a well-known Kerr nonlinearity magnitude and sign at $1550$~nm. We also considered this waveguide for plotting the theoretical broadening factor in Fig.~1, where we applied the model of Eq.~(4) to the taper and waveguide sections in a cascaded way. Hereto, we used $\alpha_{\mathrm{SOI}} = 3$~dB/cm and $\gamma_{\mathrm{SOI}} = +0.3$~W$^{-1}$mm$^{-1}$ \cite{CastelloLurbe} in the waveguide section, and we assumed a scaling of these parameters inversely proportional with the waveguide width in the tapered sections. The graphene deposition on the waveguides has been described in our earlier work \cite{VanErps}. Chemical-vapor-deposition-grown graphene was transferred on the SOI chip, inducing a high linear waveguide loss of $\alpha_{\mathrm{gr \mydash on \mydash SOI}} = 1320$~dB/cm at $1550$~nm, and the layer was patterned to vary the length of the graphene sections on top of the waveguides \cite{VanErps}. The Fermi energy of the graphene sheet was about $-0.2$~eV as a result of weak unintentional doping induced by the transfer process itself \cite{Ciuk}, so that at $1550$~nm we were operating in graphene's high-loss regime above its single-photon absorption threshold.         

We carried out SPM measurements for both a positive and negative input chirp using the experimental setup shown in Fig.~2(b). The pulsed laser sources used to excite the waveguides were a diode-pumped fiber laser (Calmar FPL-02CTF-POL; Source~(1) in Fig.~2(b)) and a fiber-coupled optical parametric oscillator or OPO (APE Levante IR with a variable optical attenuator or VOA; Source~(2)) generating, respectively, Gaussian pulses at a repetition rate of $40$~MHz and sech$^2$ pulses at a repetition rate of $80$~MHz. The pulses were coupled into and out of the waveguides by means of flat-cleaved fiber probes positioned above the waveguides' grating couplers at an angle of $9^{\circ}$ with respect to the vertical axis. We ensured excitation of the Transverse Electric (TE) mode in the waveguides using a fiber-based polarization controller (PC). To monitor the input spectrum, a 99:1 coupler was inserted in the setup to split off $1\%$ of the laser power towards a fiber-coupled optical spectrum analyzer or OSA (Yokogawa AQ6370D) just in front of the incoupling fiber probe. Simultaneously with the input spectrum, the output spectrum was measured with a second fiber-coupled optical spectrum analyzer just behind the outcoupling fiber probe. Prior to the SPM measurements, we measured the pulses' temporal characteristics by means of a frequency-resolved optical gating (FROG) instrument (Coherent Solutions HR150). The actual SPM measurements consisted of simultaneously recording the $rms$ spectral widths of the input and output spectra for the waveguides with graphene lengths varying from $0$ to $250~\mu$m in steps of $50~\mu$m.

\begin{figure}
\includegraphics[width=3.5in]{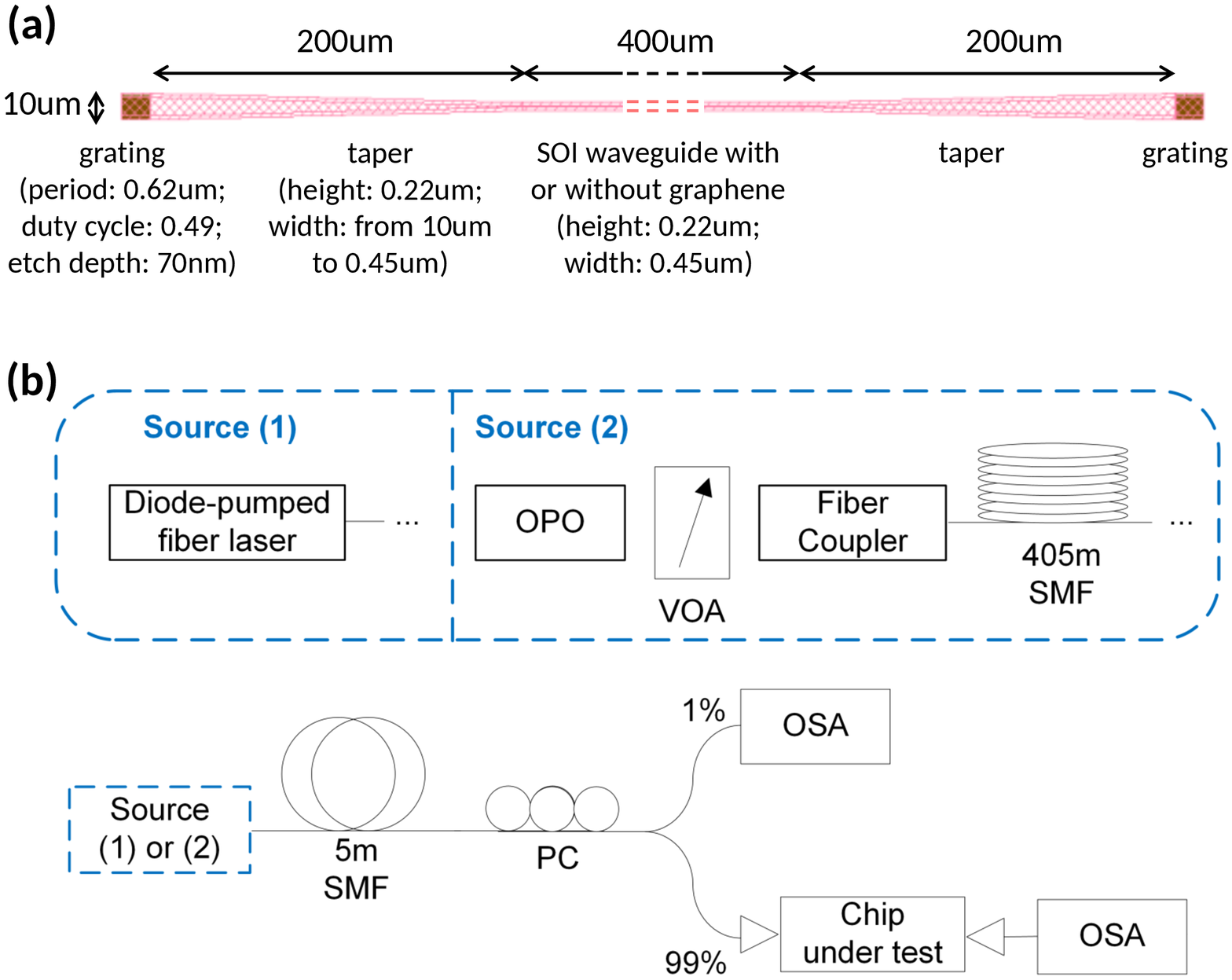}
\caption{\label{fig:ExpSetup_with_wg} (a) Outline of the employed waveguide structures. (b) Experimental setup for the SPM measurements.}
\end{figure} 

For the experiments with a positive input chirp, we employed the diode-pumped fiber laser set at an operation wavelength of $1550$~nm, with the laser pulses guided over $5$~m of G.652 single-mode fiber (SMF) towards the waveguides. The measured temporal power and phase profiles as shown in Fig.~3 yield a full-width-at-half-maximum pulse length $T_{\mathrm{FWHM},0} = 1.2$~ps and a fitted chirp parameter $C_0=0.8$. The latter was extracted from the quadratically fitted phase profile indicated as the dotted line in Fig.~3. The peak power coupled into the waveguide was set at $1.82$~W. The measured broadening factors as a function of the graphene length on top of the waveguides are depicted in Fig.~4. Since the input pulses were positively chirped, we observed for the bare SOI waveguide (graphene length of $0~\mu$m) with $\gamma_{\mathrm{SOI}} > 0$ spectral broadening as expected. The measured broadening factor lies closely to the corresponding theoretical value at $C_0=0.8$ in Fig.~1. The latter shows that the simple theory of Eq.~(4) without inclusion of two-photon absorption or free-carrier effects is an adequate model for our experiments with modest peak powers, relatively short pulse durations and short waveguide lengths. For the SOI waveguides covered with graphene, the level of spectral broadening is reduced with increasing graphene length, and for sufficiently long graphene sections even spectral narrowing is observed. We point out that, since the graphene cover layers are situated in the centre of the $400~\mu$m-long waveguide sections \cite{VanErps}, all waveguides have an uncovered SOI section at the input with $\gamma_{\mathrm{SOI}} > 0$ establishing spectral broadening. As shown in Fig.~4, this broadening is progressively counteracted by the graphene-covered sections, which consequently feature $\gamma_{\mathrm{gr \mydash on \mydash SOI}} < 0$. This observation also showcases that depositing graphene on another nonlinear material will not always enhance the nonlinear response of the material system and in some cases even annihilates its nonlinear behavior.    

\begin{figure}
\includegraphics[width=2.7in]{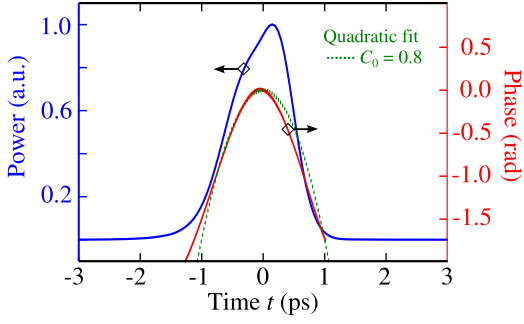}
\caption{\label{fig:Pulse_Calmar} FROG-measured temporal power profile (solid blue) and phase profile (solid red) of the input pulses with positive chirp. The dotted green line is the quadratically fitted phase.}
\end{figure}

\begin{figure}
\includegraphics[width=2.7in]{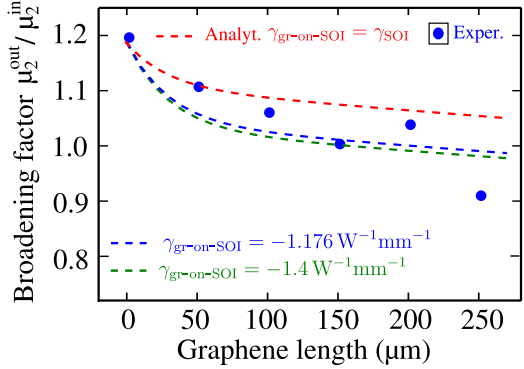}
\caption{\label{fig:Calmar_fitting} Broadening factor vs. graphene length for a positive input chirp. The analytical dashed curves serve for fitting.}
\end{figure}  

The latter finding was also confirmed in our experiments with a negative input chirp. Hereto, we employed the OPO source tuned to a wavelength of $1553$~nm. The pulses at the free-space output of the OPO are slightly positively chirped, but when coupled into a G.652 fiber with an appropriately chosen length, a negative pulse chirp can be obtained at the output of the fiber. Indeed, during propagation in the fiber, the pulses initially experience spectral broadening due to SPM effects in the fiber. At the same time, they are compressed in time due to the negative group velocity dispersion ($\beta_2 < 0$) of G.652 single-mode fiber in the telecom window around $1550$~nm. As such, after having traveled a certain propagation distance, the pulses become transform-limited ($C=0$), and beyond this point, they acquire a negative chirp induced by the fiber's negative $\beta_2$ \cite{Agrawal}. Based on numerical simulations, we found that $6$~ps-long OPO pulses with a peak power of $12$~W at the fiber input would reach this regime of negative chirp at a propagation distance of around $410$~m. Fig.~5 shows both the measured and numerically simulated temporal power and phase profiles at the output of a $410$~m-long fiber, as well as the quadratic fit of the experimental phase. The pulse length and chirp parameter are found to be $T_{\mathrm{FWHM},0} = 3$~ps and $C_0=-0.25$, respectively. The peak power that was coupled into the chip was set at $1.677$~W. Fig.~6 shows the measured broadening factors as a function of the graphene length on top of the waveguides. In view of the negative input chirp, we observed for the bare waveguide with $\gamma_{\mathrm{SOI}} > 0$ spectral narrowing, with again a very good match between the measured spectral narrowing and the corresponding theoretical value at $C_0=-0.25$ in Fig.~1. For the waveguides with graphene on top, we found that this narrowing is progressively counteracted by the graphene-covered sections exhibiting $\gamma_{\mathrm{gr \mydash on \mydash SOI}} < 0$. 

\begin{figure}
\includegraphics[width=2.7in]{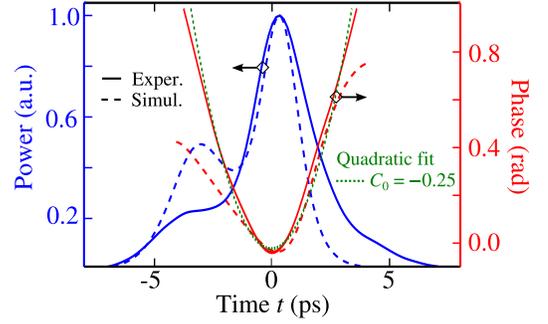}
\caption{\label{fig:Pulse_OPO} FROG-measured (solid lines) and numerically simulated (dashed lines) temporal power profile (blue) and phase profile (red) of the input pulses with negative chirp. The dotted green line is the quadratically fitted phase.}
\end{figure}

\begin{figure}
\includegraphics[width=2.7in]{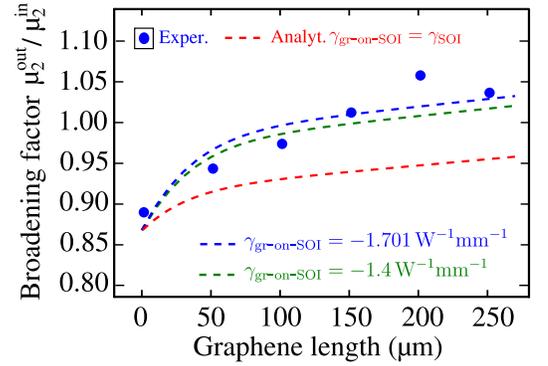}
\caption{\label{fig:OPO_fitting} Broadening factor vs. graphene length for a negative input chirp. The analytical dashed curves serve for fitting.}
\end{figure}

Besides extracting the sign of $\gamma_{\mathrm{gr \mydash on \mydash SOI}}$ in the graphene-covered waveguide sections, we also evaluated its magnitude by fitting our model with the data points of Figs.~4 and 6, while using only $\gamma_{0,\mathrm{gr \mydash on \mydash SOI}}$ as a fitting parameter. In this case, we used the more generalized model of Eq.~(2) rather than Eq.~(4), and again applied it to both the taper and waveguide sections in a cascaded way. Optimal fitting is achieved by minimizing the fitting error $\sigma=\sqrt{(1/N)\sum_{i=1}^{N}(x_i-y_i)^2}$ with $N$ the number of data points, and with $x_i, y_i$ the experimental and theoretical broadening factors, respectively. For Figs.~4 and 6, the optimal fitting was obtained for two very large, yet different nonlinearity values: $\gamma_{\mathrm{gr \mydash on \mydash SOI}} = -1.176$~W$^{-1}$ mm$^{-1}$ yielding $\sigma_{min} = 0.043$, and $\gamma_{\mathrm{gr \mydash on \mydash SOI}} = -1.701$~W$^{-1}$ mm$^{-1}$ yielding $\sigma_{min} = 0.022$, respectively. However, their average, $-1.4 (\pm 0.4)$~W$^{-1}$ mm$^{-1}$ \footnote{The tolerance interval is calculated using the same formula as for the fitting error $\sigma$ but with $x_i$ the two optimum fitting values and $y_i$ their average.}, yields an equally high-quality fitting of the experimental data. Indeed, using $\gamma_{\mathrm{gr \mydash on \mydash SOI}} = -1.4$~W$^{-1}$ mm$^{-1}$ results in approximately the same fitting error as when employing the two different nonlinearity values ($\sigma=0.044$ vs. 0.043 for Fig.~4 and $\sigma=0.024$ vs. 0.022 for Fig.~6) \footnote{Also when measuring the broadening factor as a function of input peak power instead of graphene length, we extracted the same value for $\gamma_{\mathrm{gr \mydash on \mydash SOI}}$.}. In contrast, a poor fitting is obtained when assuming the graphene-covered waveguide sections to exhibit the same nonlinearity as the bare waveguide (i.e. $\gamma_{\mathrm{gr \mydash on \mydash SOI}} = \gamma_{\mathrm{SOI}}$), which showcases the strong nonlinear impact of the graphene cover layer.

\section{Extracting the Kerr nonlinearity of graphene}

From the value $-1.4$~W$^{-1}$ mm$^{-1}$ obtained for the \emph{effective} Kerr nonlinearity $\gamma_{\mathrm{gr \mydash on \mydash SOI}}$ in the graphene-covered waveguide sections, we can extract the nonlinearity of the graphene sheet itself using our earlier presented weighted-contributions formalism for hybrid waveguides \cite{Vermeulen}. Starting from the actual waveguide mode profile in the presence of the graphene top layer and the corresponding propagation characteristics, this formalism, included in Appendix, calculates the weighted nonlinearity contributions of the graphene cover layer and of the underlying silicon material to the effective nonlinearity $\gamma_{\mathrm{gr \mydash on \mydash SOI}}$. Since $\gamma_{\mathrm{gr \mydash on \mydash SOI}}$ is the parameter that we derived from the experiments, and since we can readily calculate the weighted nonlinearity contribution of the silicon waveguide, the formula in Appendix allows extracting the nonlinearity of the graphene layer itself. This is done in two steps: first, from the formula we derive that, as most of the modal power resides in the silicon material, the contribution of the silicon to $\gamma_{\mathrm{gr \mydash on \mydash SOI}} = -1.4$~W$^{-1}$ mm$^{-1}$ is given by $\gamma_{\mathrm{SOI}}$ of the bare waveguide, and that of the graphene cover layer equals $(-1.4-0.3)=-1.7$~W$^{-1}$mm$^{-1}$. Secondly, by assessing specifically the expression for the graphene contribution (i.e. the second term in the formula in Appendix), we then extract that the graphene sheet as an isolated material features a Kerr-nonlinear index $n_{2,\mathrm{gr}}$ of about $- 10^{-13}$~m$^2$/W or a susceptibility $\chi^{(3)}_\mathrm{gr}$ around $- 10^{-7}~esu$. Its magnitude is in line with earlier reported graphene nonlinearity values \cite{Hendry, Gu} but the negative sign was not accounted for in those experiments. We point out that our earlier presented calculations for the Kerr nonlinearity of graphene \cite{Cheng1, Cheng2, Cheng3}, although underestimating its absolute value, do predict a negative sign for the nonlinearity of quasi-undoped graphene at optical wavelengths, both in the ideal case of zero electron scattering and in case of considerable inter- and intra-band scattering. The calculations show a negative Kerr nonlinearity resonance at the single-photon absorption threshold, and the negative nonlinearity sign persists also far above this threshold. 

\section{Conclusion}
In conclusion, using the chirped-pulse-pumped SPM method introduced here -- a generic method that allows extracting both the magnitude and sign of the Kerr nonlinearity of any novel (2D) material \cite{Castellanos} --, we have demonstrated a negative Kerr nonlinearity for quasi-undoped graphene, whereas the magnitude of the obtained value is in line with earlier experimental reports. Our findings imply that, when using graphene to enhance the nonlinear response of a photonic device or integrated circuit with a positive nonlinearity, one needs to carefully study how the 2D material should be implemented to actually achieve a net enhancement. Our results also show that, when targeting the compensation of unwanted nonlinearities in such a device or circuit, graphene could be of use as well. Finally, graphene's negative Kerr nonlinearity can give rise to unexplored approaches for establishing Kerr interactions and even extend the scope of e.g. the ``periodic poling'' method often used for second-order nonlinearities \cite{Hum} towards third-order Kerr processes.

\vspace{\baselineskip}

\appendix*
\section{Weighted-contributions formalism}

Below we briefly summarize the basic concepts of the weighted-contributions formalism used for extracting the graphene nonlinearity from the measured effective nonlinearity; for further details about this formalism we refer to \cite{Vermeulen}. When assuming a graphene-covered silicon waveguide oriented along the $\zeta$ direction, we can write the electric field associated with a waveguide mode as proportional to $\mathbf{e}_{\mu}(x,y)e^{ik_{\mu}\zeta}$, where $\mu = p$ indicates the mode profile excited by the pump laser used; we assume the normal to the top of the waveguide to be $\mathbf{\hat{y}}$. When determining this modal field using mode solver software, we incorporate the linear optical properties of silicon (assumed lossless at the wavelength of interest), and treat the graphene layer as a thin sheet with thickness $d_{gr}$ on top of the waveguide. We employ the function $s_{\func{Si}}(x,y)$ ($s_{gr}(x,y)$) to indicate where the silicon (graphene) is situated; $s_{\func{Si}}(x,y)=1$ ($s_{gr}(x,y)=1$) where the silicon (the thin layer modeling the graphene) is present and $s_{\func{Si}}(x,y)=0$ ($s_{gr}(x,y)=0$) where the silicon (the thin layer modeling the graphene) is absent. We obtain the following formula for the effective Kerr nonlinearity measured in the self-phase modulation experiments with the graphene-covered silicon waveguides:

\begin{widetext}

\begin{eqnarray}
\gamma_{\mathrm{gr \mydash on \mydash SOI}} \nonumber
&=& \gamma^{}_{\mathrm{SOI, weighted}} + \gamma^{}_{\mathrm{gr, weighted}} \nonumber \\ 
&=& \left( 4v_{p}^{2}  I_{p}^{2} \right)^{-1} 3\epsilon _{0}\omega _{p}\chi _{\text{Si}%
}^{(3),ijkl}\int \left( e_{p}^{i}(x^{\prime },y^{\prime })\right) ^{\ast
}e_{p}^{j}(x^{\prime },y^{\prime })\left( e_{p}^{k}(x^{\prime },y^{\prime
})\right) ^{\ast }e_{p}^{l}(x^{\prime },y^{\prime })s_{\func{Si}}(x^{\prime
},y^{\prime })\;dx^{\prime }dy^{\prime }  \nonumber \\
&&+ \left(4v_{p}^{2}d_{gr} I_{p}^{2} \right)^{-1} 3i \sigma _{\text{gr}}^{(3),ijkl}\int \left(
e_{p}^{i}(x^{\prime },y^{\prime })\right) ^{\ast }e_{p}^{j}(x^{\prime },y^{\prime
})\left( e_{p}^{k}(x^{\prime },y^{\prime })\right) ^{\ast
}e_{p}^{l}(x^{\prime },y^{\prime })s_{gr}(x^{\prime },y^{\prime })\;dx^{\prime }\;dy^{\prime },  
\nonumber \\
\end{eqnarray}

\end{widetext}

\noindent
with $\gamma^{}_{\mathrm{SOI, weighted}}$ and $\gamma^{}_{\mathrm{gr, weighted}}$ the weighted nonlinearity contributions from the underlying SOI waveguide and the graphene top layer, respectively. $v_{p}$ and $I_{p}$ are scaling factors defined in \cite{Vermeulen}, $\epsilon_0$ is the dielectric permittivity, $\omega _{p}$ is the pump frequency, $\chi _{\text{Si}}^{(3),ijkl}$ represents the third-order Kerr-nonlinear susceptibility of silicon, and $\sigma _{\text{gr}}^{(3),ijkl}$ is the third-order Kerr-nonlinear conductivity of graphene. The latter can be converted to a susceptibility using $\chi = \sigma / (-i \omega \epsilon_0 d_{gr})$ \cite{Cheng1}.

\begin{acknowledgments}
This work was supported by the ERC-FP7/2007-2013 grant 336940, the EU-FET GRAPHENICS project 618086, the Hercules-stichting grant UABR/007/09, the FWO project G.A002.13N, the EU-FP7 Graphene Flagship 604391, VUB-OZR, BELSPO-IAP, and Methusalem.
\end{acknowledgments}

\end{document}